\documentclass[amssymb,aps,floatfix,floats,preprintnumbers,showpacs,twocolumn]{revtex4-1}

\usepackage{graphicx}% Include figure files
\usepackage[applemac]{inputenc} % use extended character set

\begin{document}

\title{Chiral transport and electronic correlations in surface states of HfNiSn single crystals}
\author{L. Steinke$^1$, J. J. Kistner-Morris$^{2}$, T. F. Lovorn$^3$, H. He$^1$, A. D. Hillier$^4$, P. Miao$^1$, S. Zellman$^1$, M. Klemm$^1$, M. Green$^1$, O. Gonzalez$^1$, A. H. MacDonald$^3$ and M. C. Aronson$^1$}
\affiliation{\mbox{$^1$Department of Physics and Astronomy, Texas A \& M University, College Station, TX 77845, USA}\\
$^2$ Department of Physics and Astronomy, Stony Brook University, Stony Brook, New York, USA\\
\mbox{$^3$Department of Physics and Astronomy, University of Texas, Austin, TX , USA}\\
\mbox{$^4$ ISIS Facility, STFC Rutherford Appleton Laboratory, Chilton, Oxfordshire, OX11 0QX, UK}}
\date{\today}

\maketitle
\textbf{In most topological insulators, the valence and conduction band appear in reverse or inverted order compared to an equivalent insulator with isolated atoms. Here, we explore a different route towards topologically nontrivial states that may arise from metallic states present on the surface of bulk insulators without such band inversion. High-quality single crystals of HfNiSn show surface transport with weak anti-localization, consistent with a two-dimensional metallic state in the presence of strong spin-orbit coupling. Nonlinear $I(V)$ curves indicate electronic correlations related to a chiral, nonlocal transport component that is qualitatively similar to a quantum Hall edge state, yet in the absence of external magnetic fields. The correlations themselves may play a decisive role in creating an apparent topologically nontrivial state on the HfNiSn surface.}
\begin{figure}
\includegraphics[width=8.9 cm]{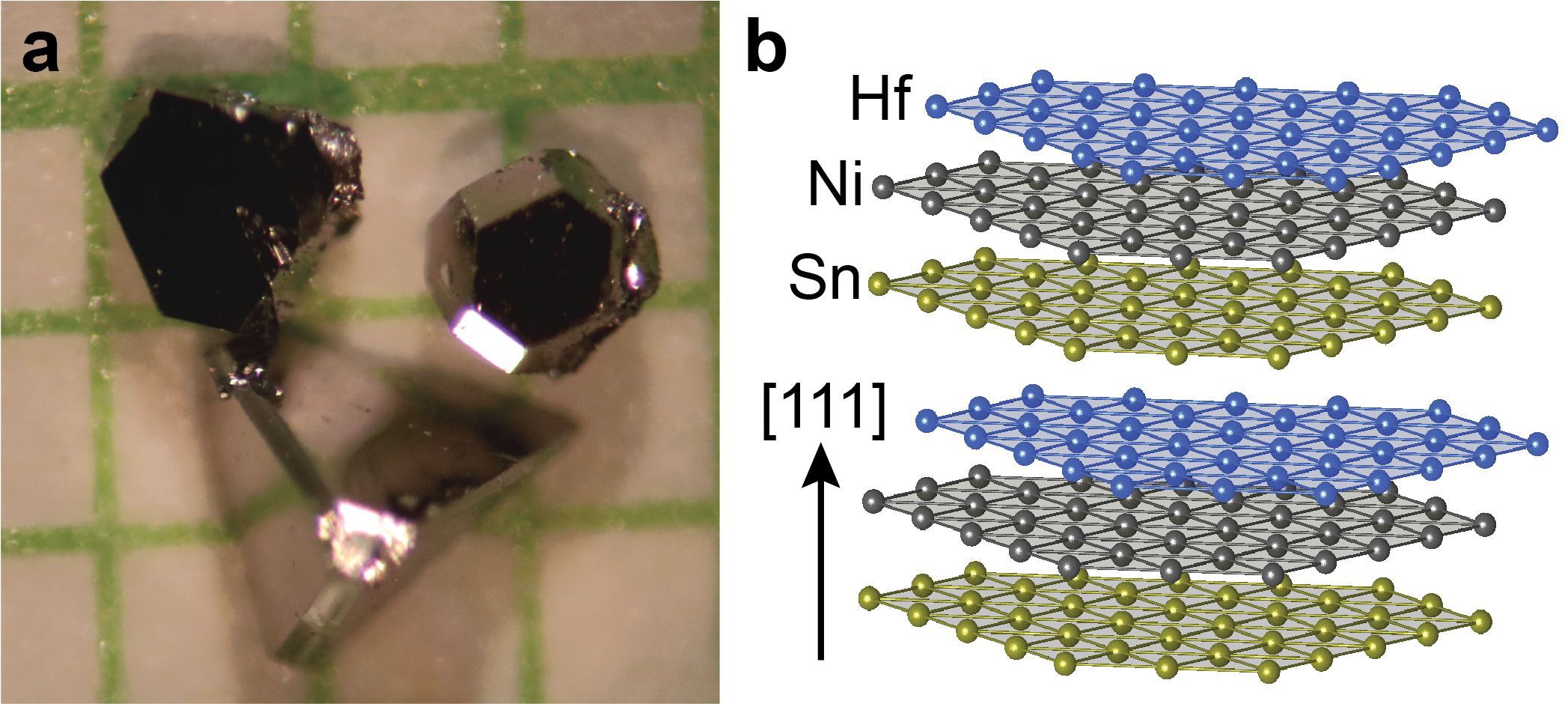}
\caption{ \textbf{HfNiSn single crystals.} \textbf{a:} photograph showing typical sizes of flux-grown HfNiSn single crystals on 1 mm graph paper. The main (111) facets have triangular or hexagonal shape.  \textbf{b:}  schematic crystal structure stretched along the [111] direction, illustrating the cyclical stacking of hexagonal layers.}
\label{Fig1}
\end{figure}

One of the most significant breakthroughs of the last decade has been the realization that a number of two-dimensional (2D) systems host topologically protected edge states with an array of remarkable properties that cannot be realized in conventional metals. Within 2D topological states such as the quantum Hall effect (QHE) systems, strong electronic correlations can induce nontrivial ground states with even more exotic phenomena like fractionally charged excitations of the fractional quantum Hall effect (FQHE). The discovery of topological insulators (TI) opens an entirely different venue, where the bulk material has a band gap, but the surface hosts gapless and 2D boundary states that are protected against backscattering by fundamental symmetries.
It is not yet known what range of states will be possible, especially if the nature and strength of correlations can be varied.

The original approach to realizing a TI was to exploit strong spin orbit coupling to invert the conduction and valence bands, first proposed for the two dimensional states of a HgTe quantum well heterostructure~\cite{Konig2007, Bernevig2006}, and subsequently for three dimensional states, as found in systems like Bi$_{1-x}$Sb$_x$ alloys~\cite{Fu2007,Hsieh2008}, Bi$_2$Se$_3$~\cite{Xia2009} and Bi$_2$Te$_3$~\cite{Chen2009}. In addition, electronic correlations can play a decisive role in 3D systems such as SmB$_6$, where the topological state is induced by the Kondo effect, and so the topological surface state cannot be understood on the basis of the single electron band structure~\cite{SmB6}.
We demonstrate here an entirely different route to the realization of a correlated two-dimensional electron gas. These topologically nontrivial states reside on the surface of a bulk insulator where band inversion is absent. This discovery is a significant advance, with potential technological relevance, since the electron gas is conveniently accessible on the surface of a bulk crystal, facilitating the manipulation of the quantum state without the need for an external magnetic field.

\begin{figure}
\includegraphics[width=8.9 cm]{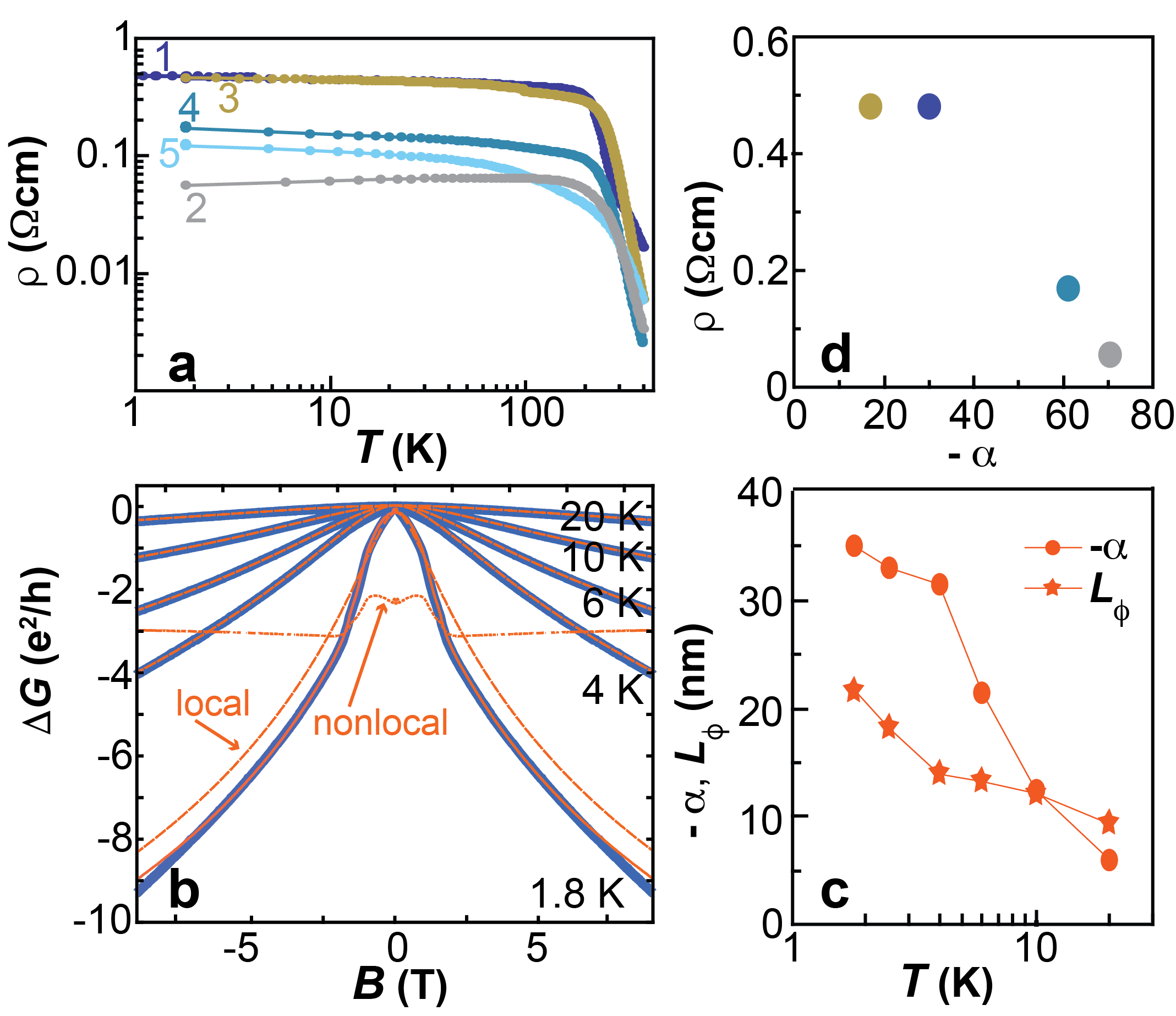}
\caption{\textbf{Crossover from 3D to 2D transport at low T.} {\textbf a}: $T$-dependence of the resistivity $\rho$ for different samples indicated by different colors.
The activated $T$-dependence is cut off below $\approx$ 200 K, where $\rho$ saturates or shows a weaker $T$-dependence consistent with variable-range hopping (cf.~\cite{sup}).
{\textbf b}: magnetoconductance $\Delta G(B) = R(B)^{-1}-R(0)^{-1}$ of sample 1 (dark blue plot in panel {\textbf a}), fitted with the HLN-expression for 2D weak anti-localization (orange curves). At 1.8 K, the $B$-dependent conductance is well described as a sum of local (WAL) and nonlocal transport components (cf. Fig.~\ref{Fig5}b).
{\textbf c}: $T$-dependence of the fit parameters $-\alpha$ and $L_\phi$ for sample 1.
{\textbf d}: saturation value of $\rho$ vs. $\alpha$ at 4 K for different samples indicated by colors in {\textbf a}.} \label{Fig2}
\end{figure}

To experimentally realize this scenario, we take advantage of the existing organizational scheme that has been established for insulating half-Heusler compounds~\cite{Chadov}, where density functional (DFT) calculations of the electronic structure show that band inversion can lead to topologically relevant or trivial states depending on the strength of the spin-orbit coupling. In agreement with this scheme, DFT calculations show that HfNiSn is a trivial bulk insulator with an indirect gap of $\simeq$ 0.3 eV, confirmed by our measurements \cite{sup}. In contrast, the low-temperature transport properties are dominated by metallic surface states.~\cite{Ahilan}.
Fig.~\ref{Fig2}a shows the $T$-dependent resistivity obtained from four-probe Van der Pauw measurements placed on a exposed 111 facet \cite{VdP}. The thermally activated transport in the bulk insulator is cut off at low temperatures $T < 200$ K, with a significantly weaker $T$-dependence or saturation of the resistivity at a value that can vary significantly among samples (Fig.~\ref{Fig2}a, b). Definitive evidence that two-dimensional electronic states with pronounced quantum character are responsible for this low temperature conduction comes from the field-dependence of the resistance. Fig.~\ref{Fig2}b shows that the magnetoconductance $\Delta G(B)=R^{-1}(B)-R^{-1}(B = 0)$ is enhanced at low temperatures and clearly deviates from a classical parabolic $B$-dependence. The $B$-dependence is well described by the Hikami-Larkin-Nagaoka (HLN) expression for weak anti-localization (Fig.~\ref{Fig2}b), a quantum interference phenomenon observed in 2D conductors in the presence of spin-orbit coupling~\cite{HLN}:
\begin{equation}
\Delta \sigma = \alpha \frac{e^2}{h}(\psi(1/2 +B_{\phi}/B)-\ln(B_{\phi}/B)),
\label{eq1}
\end{equation}
where $\psi$ is the digamma function, and $\alpha$ and $B_{\phi}$ are fit parameters relating to the number of conducting 2D modes or the phase coherence length $L_{\phi} =\sqrt{\hbar/4eB_{\phi}}$, respectively. The fits show excellent agreement with the measured data between $\sim$ 20 K and 4 K, indicating that low-temperature transport in HfNiSn is predominantly two-dimensional. Angle resolved photoemission (ARPES) measurements on YPtBi \cite{Liu2016} have already established the presence of surface states in related half-Heusler compounds. The values for $L_{\phi}$ imply that the conducting states are quantum coherent over length scales as large as 10-30 nm, similar to observations of surface states of other TIs~\cite{Assaf2013}. The value of $\alpha = n/2$ in Eq.~\ref{eq1}  counts the number $n$ of conducting 2D modes or layers, indicating that $\sim$ 100 monolayers contribute to 2D conduction in HfNiSn. While a perfect 111 surface could potentially have $\alpha=1/2$,
the larger value of $\alpha$ in natural crystals of HfNiSn presumably reflects the surface topography and roughness of these facets. Fig.~\ref{Fig2}d shows that $\alpha$ sets the overall scale for the low temperature resistivity $\rho_{T\rightarrow 0}$ in HfNiSn, indicating that the resistivity is lower when the transport involves larger numbers of conducting channels acting in parallel. We conclude that surface states dominate electronic transport as $T \rightarrow 0$, hence the bulk insulating state in HfNiSn makes no appreciable contribution to the transport at low temperatures, consistent with high crystal quality.
\begin{figure}
\includegraphics[width=8.9 cm]{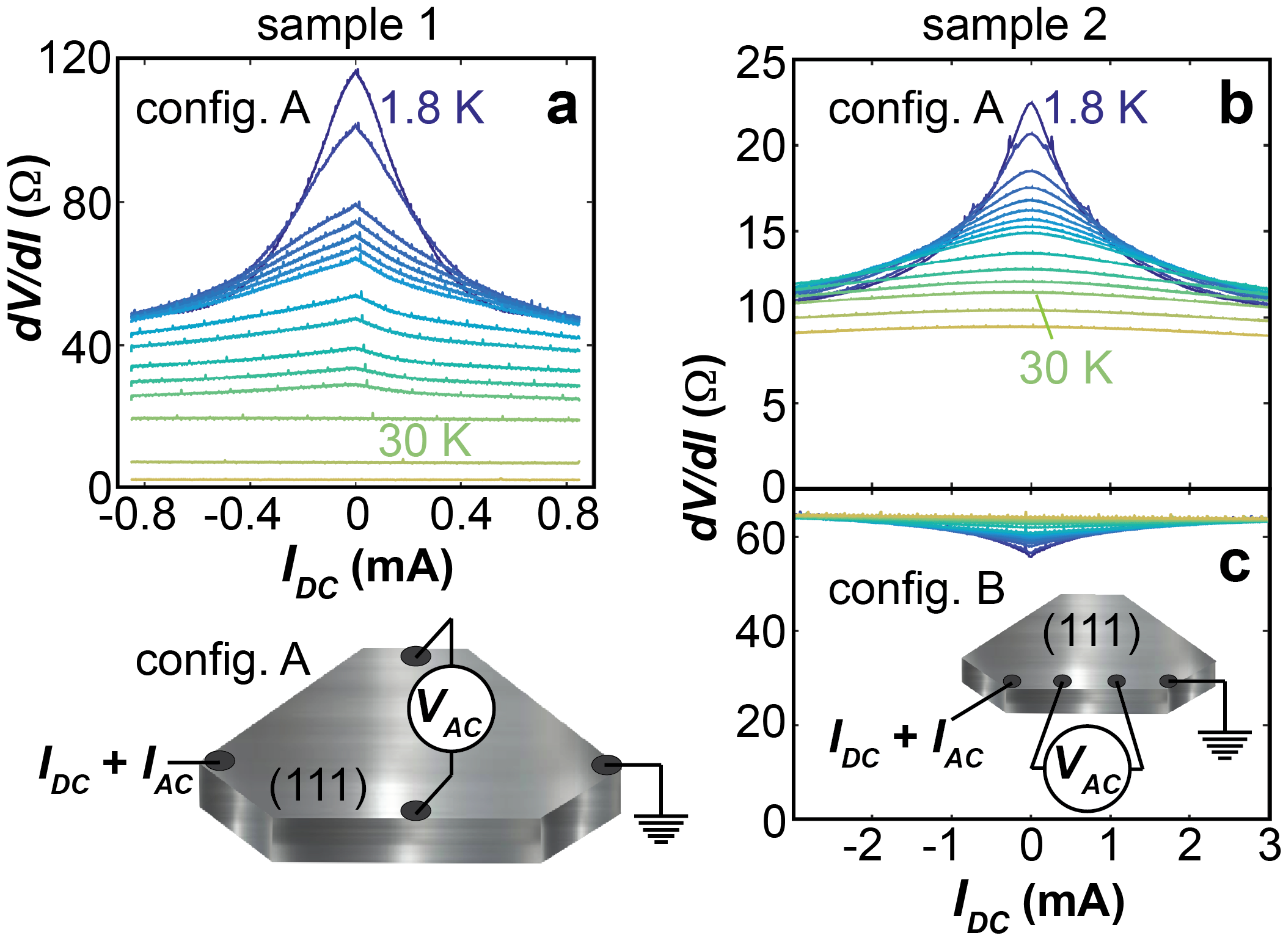}
\caption{\textbf{Nonlinear transport at low temperatures.} \textbf{a, b}: differential resistance $dV/dI$ measured between 1.8 K and 50 K in the transverse 4-point configuration B, where the voltage is measured perpendicular to the current path. {\textbf c}: $dV/dI$ measured in configuration A, where $V$ is measured along the current path.}
\label{Fig3}
\end{figure}

These surface states in HfNiSn are apparently strongly correlated, a necessary ingredient for certain topologically non trivial states, such as those found in the three-dimensional TI SmB\textsubscript{6}~\cite{SmB6} and the FQHE~\cite{FQHE1,FQHE2}.  Differential resistivity $dV/dI(I_{DC})$ measurements display pronounced nonlinearities for temperatures less than 30 K (Fig.~\ref{Fig3} a,b), establishing that the density of states for the current carrying excitations has a nonlinear energy dependence. Intriguingly, the observed nonlinearities depend strongly on the 4-point contact configuration used for the measurements, and are most pronounced when the voltage is measured perpendicular to the direction of the applied current (Fig.~\ref{Fig3} a,b). This is our first indication that the surface conduction in HfNiSn has pronounced nonlocal character. Nonlocal conduction is a signature property of low dimensional conductors, such as surface and edge states.

\begin{figure}
\includegraphics[width=8.9 cm]{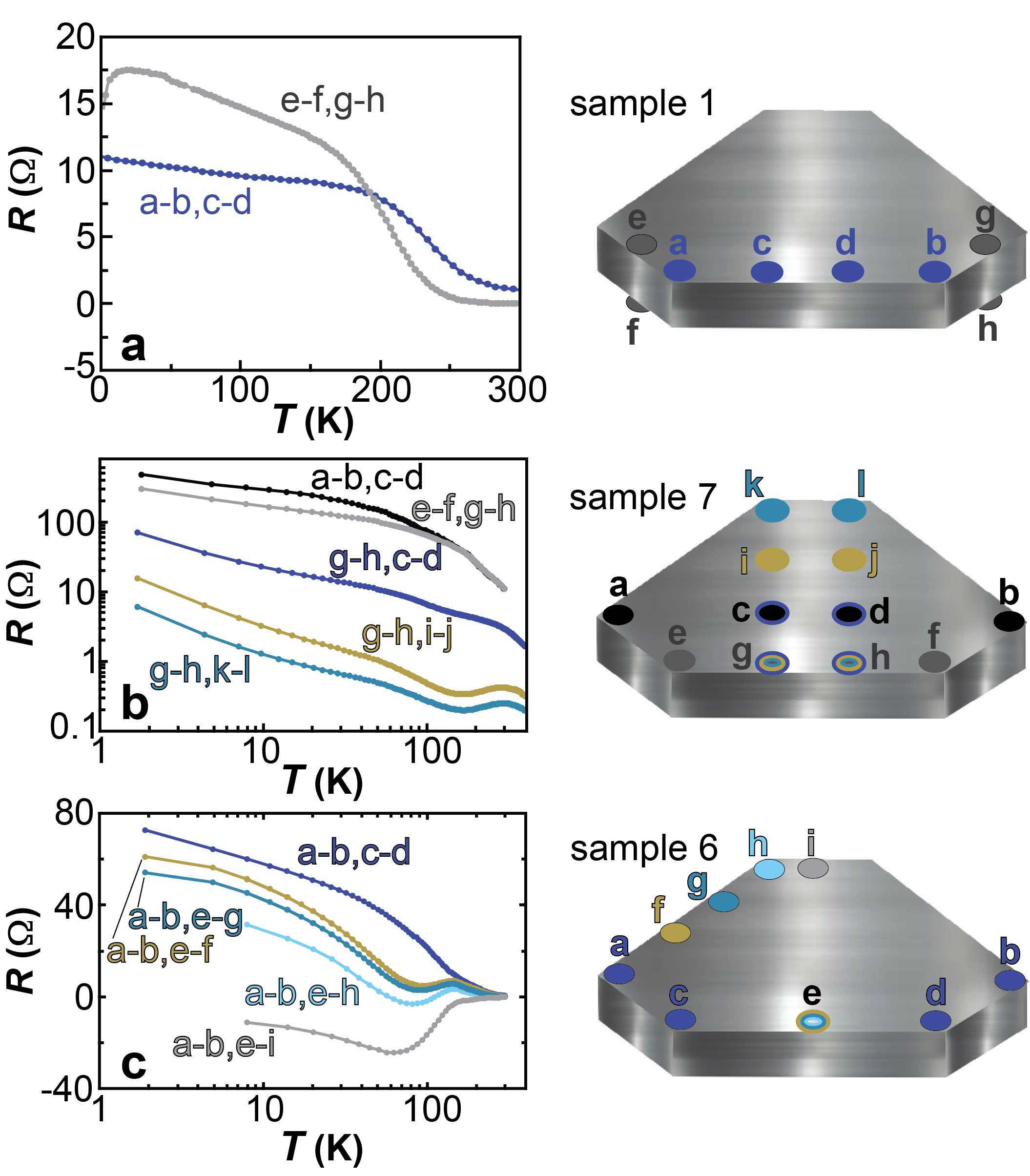}
\caption{\textbf{Local and nonlocal transport in HfNiSn, with distinct $T$-dependence.} {\textbf a}: temperature dependence of the resistance within the top $[111]$ surface (blue) and nonlocal resistance measured between contacts on top and bottom surface (gray), as illustrated in the inset. Four-point configurations are labeled by the contacts used, where the first pair are the current contacts and the second pair the voltage contacts. {\textbf b}: comparison of increasingly nonlocal voltage measurements within the top [111] facet, varying the distance from the direct current path. {\textbf c}: comparison of increasingly nonlocal voltage measurements, varying the angle between the current path and measured potential drop. Measurements in panels {\textbf a}, {\textbf b}, and {\textbf c} were performed on different samples, and in absence of an external magnetic field.}
\label{Fig4}
\end{figure}

Nonlocal transport, where current travels along preferred paths, other than the direct path between current terminals, causes potential drops that can be observed at large distances. Its presence in HfNiSn is demonstrated in various four-point contact configurations illustrated in Fig.~\ref{Fig4}.
The gray curve in Fig.~\ref{Fig4}a shows a four-probe resistance measurement where the current $I_{e-f}$ was applied between neighboring contacts $(e,f)$ on the top and bottom (111)-surfaces of a HfNiSn single crystal, separated by the sample thickness of $\sim 0.2$ mm (sample 1, Fi.~\ref{Fig4}a). If conduction were purely local, no voltage $V_{g-h}$ would be expected between a pair of contacts $(g,h)$ located at a lateral distance of $\sim 1$ mm from the current terminals, and this is indeed the case at 300 K. As the temperature is lowered below 200 K and surface conduction becomes more dominant, this nonlocal voltage $V_{g-h}$ increases markedly, and the nonlocal resistance $R_{e-f,g-h} = V_{g-h}/I_{e-f}$ even exceeds the in-plane resistance $R_{a-b,c-d}$ within a single facet (blue curve in panel a). A simple experiment where voltages are measured between contact pairs placed only in the top (111) facet (sample 7, Fig.~\ref{Fig4}b), but located at increasing distances from the direct current path ($(a,b)$ in panel b), or at increasing angles relative to the current direction (sample 6, Fig.~\ref{Fig4}c), allows us to separate the local and nonlocal voltages on the basis of their temperature dependencies. The most nonlocal contact configuration has the voltage terminals perpendicular to the current path, where zero voltage would be observed in a conventional system (gray curve in Fig.~\ref{Fig4}c). In each case, a monotonic increase in $R(T)$ is observed, and we ascribe this predominantly to the local conduction, whose temperature dependence is well described by variable range hopping~\cite{sup}. Our analysis does not indicate whether the associated impurity states reside in bulk insulating HfNiSn or in its surface. A minimum in $R(T)$ develops near 70 K, becoming more pronounced and even becoming negative as the contact configuration becomes increasingly nonlocal, showing that a second conduction mechanism associated with nonlocal transport is present.

\begin{figure}
\includegraphics[width=8.9 cm]{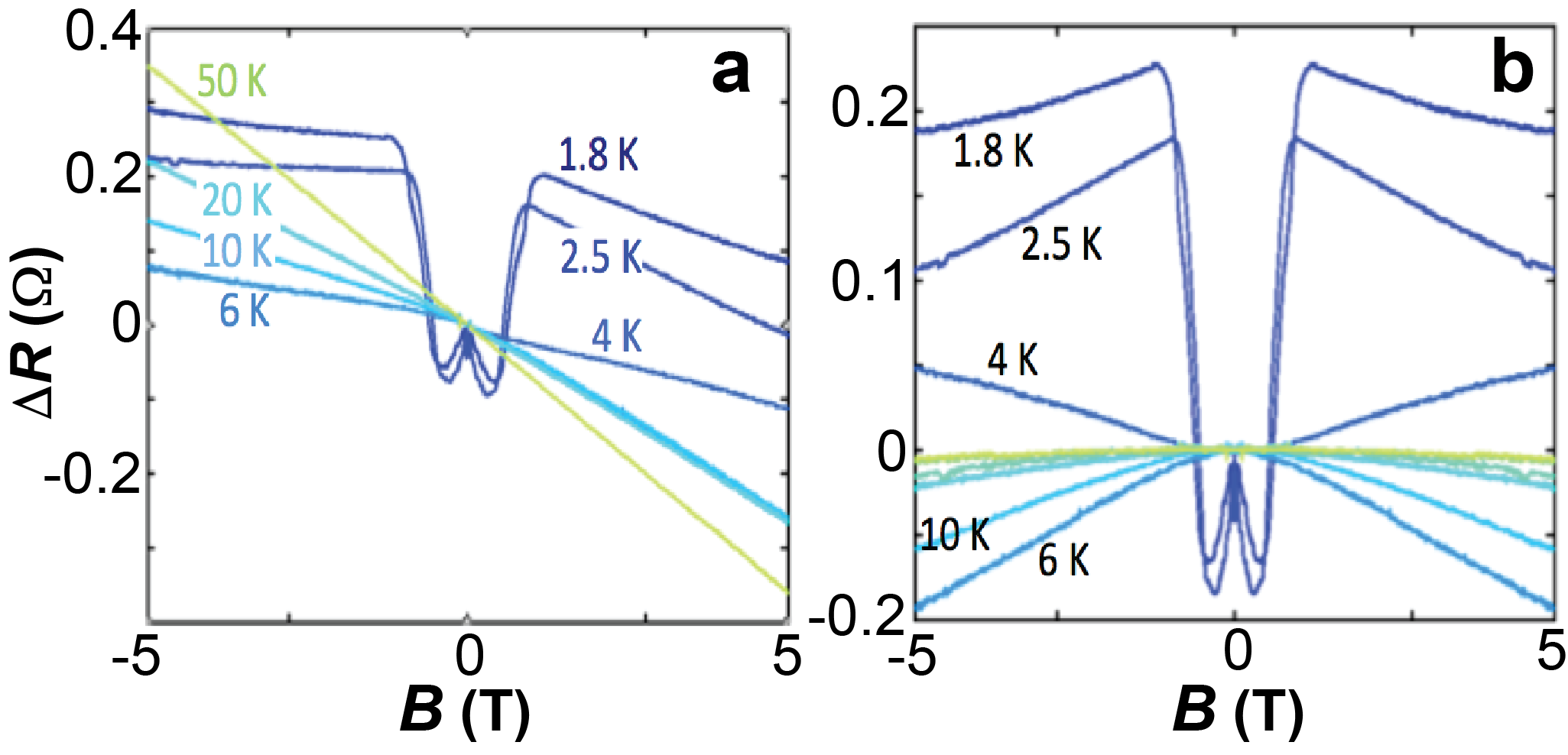}
\caption{\textbf{Nonlocal transport with characteristic $B$-dependence.} {\textbf a}: Nonlocal magnetoresistance $\Delta R = R (B) - R (B = 0)$, measured in a transverse geometry (Fig.~\ref{Fig4}c, configuration $I: a-b$, $V: e-i$), where at low $T$, a $B$-symmetric nonlocal contribution dominates over the classical Hall effect. Panel {\textbf b} shows only this even component.}
\label{Fig5}
\end{figure}
\begin{figure*}
\includegraphics[width=18.3 cm]{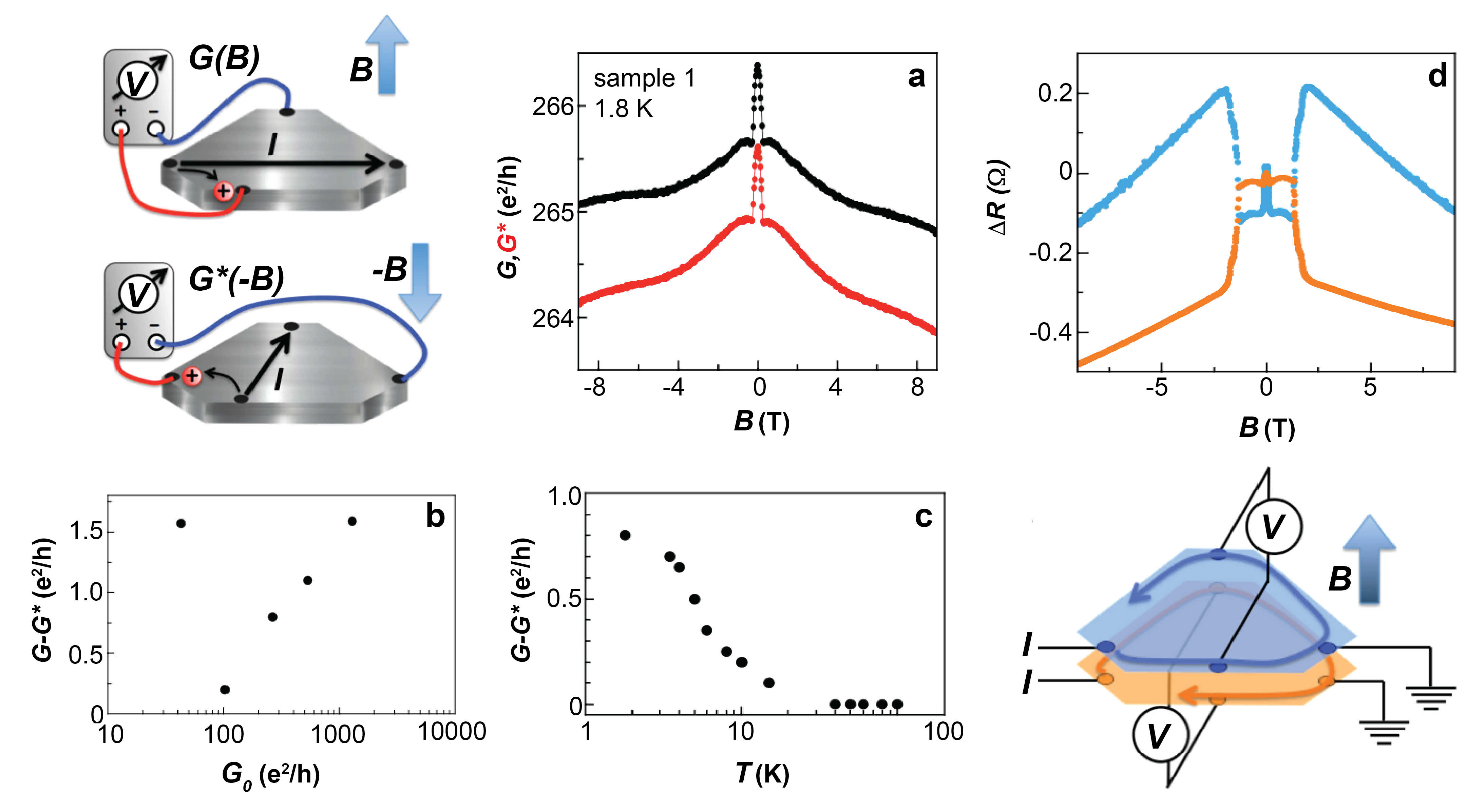}
\caption{\textbf{Evidence for Time-reversal symmetry breaking in HfNiSn.} {\textbf a}: comparison of nonlocal magnetoconductance $G_{XY}=R_{XY}^{-1}$, measured in two illustrated geometries $G_{XY}(B)$ and $G_{XY}^{\star}(-B)$ related by Onsager reciprocity. The difference $\Delta G$ between the two measurements , which indicates TRS breaking, corresponds to $\sim~1 e^2/h$. Panel {\bf b} shows $\Delta G$ for various samples, and {\bf c} shows the $T$-dependence of $\Delta G$ for sample 1. Panel {\textbf d} compares simultaneous measurements of the nonlocal MR in identical contact configurations, and on opposite (111) facets, as illustrated below.}
\label{Fig6}
\end{figure*}
The field dependencies of the local and nonlocal contributions to the conduction are also distinct, and can be studied using different four-probe configurations (Fig.~\ref{Fig5}). As we have discussed above (Fig.~\ref{Fig2}b), the field dependence of the local conductivity is explained in terms of WAL, and can be readily contrasted to that of the nonlocal magnetoresistance $\Delta R$ (MR) in Fig.~\ref{Fig5}a, which is measured with the voltage terminals placed perpendicular to the current path - a configuration typically used to measure the Hall effect. At high temperatures, the classical Hall effect dominates, with a linear field dependence and a negative slope indicating electron conduction, and a temperature dependence consistent with the freeze-out of thermally activated carriers~\cite{sup}. Below $\sim$ 50 K, the slope decreases, and the classical Hall effect is almost completely suppressed. This is the temperature regime where the nonlocal conduction begins to dominate in zero field. It is remarkable that $\Delta R$ is also symmetric in field, ruling out the presence of either a conventional or anomalous Hall effect. Clearly, the detailed field dependencies of the local and nonlocal conductivities are not the same. This disparity becomes all the more pronounced as the temperature is lowered, where the nonlocal MR shows a wholly different temperature dependence, with negative MR from 20 K to 6 K, and positive MR below 4 K. At 1.8 K, the nonlocal transport becomes so dominant that it can even be detected with the contact configuration that leads to the most local conduction (Fig.~\ref{Fig2}b). Here, the departures from WAL correspond very nicely to the MR detected in the most nonlocal configurations (Fig.~\ref{Fig5}b).

The dramatically different field and temperature dependencies for local and nonlocal transport indicate different transport mechanisms for the two conductance components. At low temperatures, the local transport is associated with the 2D metallic surface state of the HfNiSn crystals, and is consistent with a homogenous current distribution, as in a simple conductor. The nonlocal conductance is of a similar magnitude at the lowest temperatures, and is enhanced when the voltage contacts are moved to the edge of the crystal facet (configurations a-b,c-d (black) vs. e-f,g-h (gray) in Fig.~\ref{Fig4}b).
This implies that nonlocal charges travel along the edges of the main (111) facets. We consider it to be less likely that preferential current paths are caused by crystal defects or metallic inclusions, since nonlocal transport characteristics are reproducible among multiple crystals, and only seem to depend on the relative placement of current and voltage contacts (cf. Fig.~\ref{Fig4}).
A larger current density at the edge of the (111) facet raises the interesting possibility that the nonlocal conduction in HfNiSn is associated with edge states. 

The presence of edge states can result in the macroscopic breaking of time reversal symmetry (TRS), as in quantum Hall systems~\cite{QHE_edges} and in the (quantum) anomalous Hall effect~\cite{AHE,QAHE}. Magnetic susceptibility measurements show that HfNiSn is strongly diamagnetic, and $\mu$SR measurements find no evidence for static magnetic fields \cite{sup}. Thus, there is no experimental evidence for bulk magnetic order in HfNiSn, a result that is supported by its absence in electronic structure calculations. Consequently, the violation of Onsager reciprocity is experimental proof of the breaking of TRS, and if present in HfNiSn, this would be evidence that these edge states are chiral. Testing Onsager reciprocity requires measurements of the nonlocal conductances $G$ and $G^{\star}$~\cite{OC} using two different four-point configurations, where current and voltage terminals are swapped between measurements as illustrated in Fig.~\ref{Fig6}.

The measured results will be identical at $B = 0$, if TRS is preserved \cite{OC}. If TRS is explicitly broken by an applied magnetic field, the resulting deflection of charge (sketched for positive carriers in Fig.~\ref{Fig6}, top left inset) causes a difference between $G$ and $G^{\star}$ that is accounted for by comparing $G(+B)$ to $G^{\star} (-B)$ at the exact opposite field. Fig.~\ref{Fig6}a compares measurements of $G(B)$ and $G^{\star}(-B)$ at 1.8 K, showing a clear separation between the two curves that is present up to our maximum field of 9 T. The observed difference $\Delta G = G(B) - G^{\star}(-B)$ indicates that there is an inherent breaking of time reversal symmetry in the surface states of HfNiSn that does not depend on the presence of a magnetic field.

The magnitude of $\Delta G$ approaches one conductance quantum $e^2 / h$, and this order of magnitude has been verified among a range of different samples having values of $G_0 = G(B=0)$ that differ by as much as three orders of magnitude (Fig.~\ref{Fig6}b). The temperature dependence of $\Delta G$ (Fig.~\ref{Fig6}c) shows that the TRS breaking only occurs below $\sim$ 30 K, the temperature where electronic correlations first appear (Fig.~\ref{Fig3}). It seems likely that correlations that are not strong enough to lead to order are the underlying cause of the TRS breaking observed in the nonlocal transport in HfNiSn. 

Is the correlated surface state observable on macroscopic length scales? Fig.~\ref{Fig6}d compares measurements performed simultaneously in identical contact configurations, but on opposite facets. Amazingly, these measurements show exactly opposite field-dependent responses. The likely interpretation is that the chiral edge states implied by our measurements have opposite handedness on the top and bottom facets, a result that may be a consequence of the underlying noncentrosymmetric crystal structure of HfNiSn (Fig.~\ref{Fig1}), where the cyclical stacking of hexagonal (111) layers renders opposite (111) facets inequivalent. 

%\section{Conclusion}
We conclude that high-quality single crystals of HfNiSn host a remarkably rich two-dimensional state on their (111) surfaces. While the bulk crystal is an insulator at low temperatures, the electronic system on its surface is strongly correlated and shows two conduction mechanisms characterized by local and nonlocal transport. The electronic correlations are concomitant with time-reversal symmetry breaking of the nonlocal transport component, in the complete absence of external magnetic fields or sample magnetism. This suggests a chiral transport mechanism. Chiral, nonlocal transport is a hallmark feature of 2D topological systems with protected edge states, like QHE \cite{QHE_edges}, quantum spin Hall effect \cite{Konig2007, Bernevig2006}, and quantum anomalous Hall effect systems~\cite{QAHE}. Our experiments indicate that such a state may be naturally present on the HfNiSn surface. In quantum Hall systems, the interplay between magnetic fields and electronic interactions can lead to even more exotic phenomena like the fractional quantum Hall effect. In fact, there are remarkable similarities between transport in a fractional quantum Hall system and our samples: at magnetic fields outside a pure FQHE state, where current is carried by both the bulk 2D system (local transport) and FQHE edge states (nonlocal transport), nonlinear $dV/dI$ characteristics are predominantly seen in the nonlocal component dominated by edge states~\cite{Wang1992}. Even the edge state chirality, reflected in the asymmetric $dV/dI$-curve in~\cite{Wang1992}, appears to show in our $dV/dI$-measurement as well. We note that sample 1 shows a stronger asymmetry, consistent with a larger chiral component $\Delta G\approx 1\,{\rm e^2/h}$, compared to only $\approx 0.2\,{\rm e^2/h}$ in sample 2 with the more symmetric $dV/dI$-characteristics.

These results suggest an alternative route to 2D topological states in HfNiSn and possibly in related compounds, without the need for MBE-grown thin films, free-standing monolayers, or high magnetic fields. Their natural presence on the surface of a bulk single crystal could make such topological states attractive candidates for electronic applications, as the quantum state is easily accessible to contacting or manipulation by gate electrodes near the surface. While the origin of the topological state in HfNiSn is still unclear, it is highly probable that it is linked to electronic correlations, and cannot be predicted from a single electron approach. Understanding the possible link between crystal symmetry, surface states, and electronic correlations poses an important challenge to theory, to reveal the mechanisms leading to this exotic topological state.

\section*{Data availability}
The datasets generated during and/or analyzed during the current study are available from the corresponding author on reasonable request.

\section*{Acknowledgement}
The authors would like to thank A. Nevidomskyy, X. Quian, S. Yang, J. Denlinger, A. Finkelstein, V. Zyuzin, M. Foster, G. Gervais, and M. Grayson for illuminating discussions. This work was supported by the Army Research Office.

\end{document}